\documentclass[chi_draft]{sigchi}

\CopyrightYear{2016}
\setcopyright{acmlicensed}
\doi{http://dx.doi.org/10.475/123_4}
\isbn{123-4567-24-567/08/06}
\conferenceinfo{CHI'16,}{May 07--12, 2016, San Jose, CA, USA}
\acmPrice{\$15.00}

\usepackage{balance}       
\usepackage{graphics}      
\usepackage[T1]{fontenc}   
\usepackage{txfonts}
\usepackage{mathptmx}
\usepackage[pdflang={en-US},pdftex]{hyperref}
\usepackage{color}
\usepackage{booktabs}
\usepackage{textcomp}
\usepackage{cuted}
\usepackage{capt-of}
\usepackage{wrapfig}

\usepackage{microtype}        
\usepackage{ccicons}          

\def\plaintitle{Real-Time Lip Sync for Live 2D Animation}

\def\emptyauthor{}
\def\plainkeywords{Cartoon Lip Sync; Live animation; Machine learning.}

\makeatletter
\def\url@leostyle{%
  \@ifundefined{selectfont}{
    \def\UrlFont{\sf}
  }{
    \def\UrlFont{\small\bf\ttfamily}
  }}
\makeatother
\def\pprw{8.5in}
\def\pprh{11in}

\definecolor{linkColor}{RGB}{6,125,233}
\hypersetup{%
  pdftitle={\plaintitle},
  pdfauthor={\emptyauthor},
  pdfkeywords={\plainkeywords},
  pdfdisplaydoctitle=true, 
  bookmarksnumbered,
  pdfstartview={FitH},
  colorlinks,
  citecolor=black,
  filecolor=black,
  linkcolor=black,
  urlcolor=linkColor,
  breaklinks=true,
  hypertexnames=false
}

\newcommand{\full}[0]{\emph{Ours}}
\newcommand{\some}[0]{\emph{Ours2/3}}
\newcommand{\noaug}[0]{\emph{OursNoAug}}
\newcommand{\past}[0]{\emph{OursPast}}

\newcommand{\chon}[0]{\emph{ChOn}}
\newcommand{\choff}[0]{\emph{ChOff}}
\newcommand{\tboff}[0]{\emph{TBOff}}

\newif\ifsubmit
\submittrue
\ifsubmit
    \newcommand{\todo}[1]{}
    \newcommand{\figtodo}[1]{}
    \newcommand{\da}[1]{}
    \newcommand{\wil}[1]{}
\else
    \newcommand{\todo}[1]{{\color{red}{[}\emph{#1}{]}}}
    \newcommand{\figtodo}[1]{{\textcolor{red}{{\bfi (MAKE FIGURE)} #1}}}
    \newcommand{\da}[1]{\textbf{\textcolor{blue}{DA: #1}}}
    \newcommand{\wil}[1]{\textbf{\textcolor[rgb]{0, 1, 0}{WL: #1}}}
\fi

\makeatletter
\def\@copyrightspace{\relax}
\makeatother

\begin{document}

\title{\plaintitle}

\numberofauthors{3}
\author{%
  \alignauthor{Deepali Aneja\\
    \affaddr{University of Washington}\\
    \email{deepalia@cs.washington.edu}}\\
  \alignauthor{Wilmot Li\\
    \affaddr{Adobe Research}\\
    \email{wilmotli@adobe.com}}\\
}
\maketitle

\begin{strip}\centering
\includegraphics[width=\textwidth]{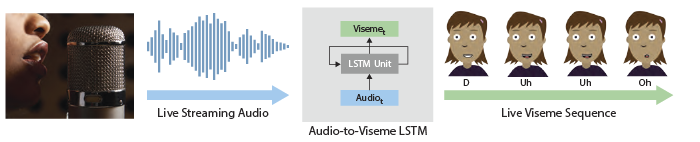}
\captionof{figure}{Real-Time Lip Sync. Our deep learning approach uses an LSTM to convert live streaming audio to discrete visemes for 2D characters.
\label{fig:feature-graphic}}
\end{strip}

\begin{abstract}
The emergence of commercial tools for real-time performance-based
2D animation has enabled 2D characters to appear on live broadcasts
and streaming platforms. A key requirement for live animation 
is fast and accurate lip sync that
allows characters to respond naturally to other actors or the
audience through the voice of a human performer.
In this work, we present a deep learning based interactive system that automatically
generates live lip sync for layered 2D characters using a Long Short
Term Memory (LSTM) model.
Our system takes streaming audio as input and produces viseme sequences 
with less than 200ms of latency (including processing time). 
Our contributions include specific design decisions for our feature
definition and LSTM configuration that provide a small but useful amount
of lookahead to produce accurate lip sync. We also describe a data augmentation 
procedure that allows us to achieve good results with a very small
amount of hand-animated training data (13-20 minutes). 
Extensive human judgement experiments show that our results are preferred
over several competing methods, including those that only support offline
(non-live) processing. Video summary and supplementary results at GitHub link: \url{https://github.com/deepalianeja/CharacterLipSync2D}
\end{abstract}

\category{I.3.3.}{Computer Graphics}{Animation}


\section{Introduction}

For decades, 2D animation has been a popular storytelling medium across many
domains, including entertainment, advertising and education. Traditional
workflows for creating such animations are highly labor-intensive; animators
either draw every frame by hand (as in classical animation) or manually specify
keyframes and motion curves that define how characters and objects move.  %
However, \emph{live 2D animation} has recently emerged as a powerful new way to
communicate and convey ideas with animated characters. In live animation, human
performers control cartoon characters in real-time, allowing them to interact
and improvise directly with other actors and the audience. Recent
examples from major studios include Stephen Colbert interviewing cartoon guests on The Late
Show~\cite{lateShow}, Homer answering phone-in questions from viewers during a segment
of The Simpsons~\cite{simpsons}, Archer talking to a live audience at ComicCon~\cite{archer},
and the stars of animated shows (e.g., Disney's Star vs. The Forces of Evil, My Little Pony, cartoon Mr. Bean) hosting live chat
sessions with their fans on YouTube and Facebook Live. In addition to these
big budget, high-profile use cases, many independent podcasters and game
streamers have started using live animated 2D avatars in their shows. 

Enabling live animation requires a system that can capture the performance of a
human actor and map it to corresponding animation events in real time.  For
example, Adobe Character Animator (Ch) --- the predominant live 2D animation tool
--- uses face tracking to translate a performer's facial expressions to a cartoon
character and keyboard shortcuts to enable explicit triggering of animated
actions, like hand gestures or costume changes.  While such features give
performers expressive control over the animation, the dominant component of
almost every live animation performance is speech; in all the examples mentioned
above, live animated characters spend most of their time talking with other
actors or the audience. As a result, the most critical type of performance-to-animation mapping for  live animation is \emph{lip sync} --- transforming an
actor's speech into corresponding mouth movements in the animated character.
Convincing lip sync allows the character to embody the live performance, while
poor lip sync breaks the illusion of characters as live participants. In this
work, we focus on the specific problem of creating high-quality lip sync for
live 2D animation.

Lip sync for 2D animation is typically done by first creating a discrete set of
mouth shapes (visemes) that map to individual units of speech for each
character. To make a character talk, animators choose a timed sequence of
visemes based on the corresponding speech. Note that this process differs from
lip sync for 3D characters. While such characters often have predefined blend
shapes for common mouth poses that correspond to visemes, the animation process
involves smooth interpolation between blend shapes, which moves the mouth in a
continuous fashion. The discrete nature of 2D lip sync gives rise to some unique
challenges. First, 2D animators have a constrained palette with which to produce
convincing mouth motions. While 3D animators can slightly modify the mouth shape
to produce subtle variations, 2D animators almost always restrict themselves to
the predefined viseme set, since it requires significantly more work to author
new viseme variations. Thus, choosing the appropriate viseme for each sound in
the speech is a vital task. Furthermore, the lack of continuous mouth motion
means that the timing of transitions from one viseme to the next is critical to
the perception of the lip sync. In particular, missing or extraneous transitions can make the animation look out of sync with the speech. Given these
challenges, it is not surprising that lip sync accounts for a significant
fraction of the overall production time for many 2D animations. In
discussions with professional animators, they estimated five to seven hours of
work per minute of speech to hand-author viseme sequences.

Of course, manual lip sync is not a viable option for our target application of
live animation. For live settings, we need a method that automatically generates
viseme sequences based on input speech. Achieving this goal requires addressing
a few unique challenges. First, since live interactive performances do not
strictly follow a predefined script, the method does not have access to an
accurate transcript of the speech.  Moreover, live animation requires
real-time performance with very low latency, which precludes the use of accurate
speech-to-text algorithms (which typically have a latency of several seconds)
in the processing pipeline. More generally, the low-latency requirement prevents
the use of any appreciable ``lookahead'' to determine the right viseme for a
given portion of the speech.  Finally, since there is no possibility to manually
refine the results after the fact, the automatic lip sync must be robust.

In this work, we propose a new approach for generating live 2D lip sync. To
address the challenges noted above, we present a real-time processing pipeline
that leverages a simple Long Short Term Memory (LSTM) \cite{hochreiter1997long} model to convert streaming audio input into a
corresponding viseme sequence at 24fps with less than 200ms latency (see
Figure~\ref{fig:feature-graphic}). While our system largely relies on an existing
architecture, one of our contributions is in identifying the appropriate feature
representation and network configuration to achieve state-of-the-art results for
live 2D lip sync.  Another key contribution is our method for collecting
training data for the model. As noted above, obtaining hand-authored lip sync
data for training is expensive and time-consuming. Moreover, when creating lip
sync, animators make stylistic decisions about the specific choice of visemes
and the timing and number of transitions. As a result, training a single
``general-purpose'' model is unlikely to be sufficient for most applications.
Instead, we present a technique for augmenting hand-authored training data 
through the use of audio time warping \cite{berndt1994using}. In particular, we ask
animators to lip sync sentences from the TIMIT~\cite{timit} dataset that have
been recorded by multiple different speakers. After providing the lip sync for
just one speaker, we warp the other TIMIT recordings of the same sentence to
match the timing of the first speaker, which allows us to reuse the same lip
sync result on multiple different input audio streams. 

We ran human preference experiments to compare the quality of our method to
several baselines, including both \emph{offline} (i.e., non-live) and 
\emph{online} automatic lip sync from two commercial 2D animation tools. Our
results were consistently preferred over all of these baselines, including the
offline methods that have access to the entire waveform. We also analyzed the tradeoff between lip sync
quality and the amount of training data and found that our data augmentation
method significantly improves the output of the model. The experiments indicate
that we can produce reasonable results with as little as 13-15 minutes of
hand-authored lip sync data. Finally, we report preliminary findings that
suggest our model is able to learn different lip sync styles based on the
training data.

\section{Related work}
There is a large body of previous research that analyzes speech input to
generate structured output, like animation data or text. Here we summarize the
most relevant areas of related work.

\subsection*{Speech-Based Animation}

Many efforts focus on the problem of automatic lip sync, also known as
speech-based animation of digital characters.  
Most solutions fall into one of three general categories: procedural techniques
that use expert rules to convert speech into animation; database (or unit
selection) approaches that repurpose previously captured motion segments or
video clips to visualize new speech input; and model-driven methods that learn
generative models for producing lip sync from speech. 

While some of these approaches achieve impressive results, the vast majority
rely on accurate text or phone labels for the input speech. 
For example, the recent JALI system by Edwards et al.~\shortcite{edwards2016jali} takes a
transcript of the speech as part of the input, and many other methods
represent speech explicitly as a sequence of phones~\cite{kim2015decision,fan2015photo,cohen1993modeling,massaro1999picture,taylor2012dynamic, taylor2017deep,lee2002audio}.
A text or phone-based representation is beneficial because it abstracts away
many idiosyncratic characteristics of the input audio, but generating an accurate
transcript or reliable phone labels is very difficult to do in real-time, with
small enough latency to support live animation applications. The most
responsive real-time speech-to-text (STT) techniques typically require several
seconds of lookahead and processing time~\cite{SpeechRecognitionBlog}, which is
clearly unacceptable for live interactions with animated characters. 
Our approach foregoes an explicit translation into phones and learns a direct
mapping between low-level audio features and output visemes that can be applied
in real-time with less than 200ms latency.

Another unique aspect of our problem setting is that we focus on generating
discrete viseme sequences. In contrast, most previous lip sync techniques aim to
produce ``realistic'' animations where the mouth moves smoothly between poses. 
Some of these methods target rigged 3D characters or meshes with predefined
mouth
blendshapes that correspond to speech sounds~\cite{xu2013practical,karras2017audio,taylor2012dynamic,edwards2016jali,mattheyses2015audiovisual,suwajanakorn2017synthesizing}, while
others generate 2D motion trajectories that can be used to deform facial images
to produce continuous mouth motions~\cite{cao2005expressive,brand1999voice}. As noted earlier, discrete 2D lip sync is not designed to be smooth or
realistic. Animators use artistic license to create viseme sequences that capture
the essence of the input speech. Operationalizing this artistic process requires
different techniques and different training data than previous lip sync methods
that aim to generate realistic, continuous mouth motions. 
In the domain of discrete 2D lip sync, one relevant recent system is
Voice Animator \cite{furukawa2017voice}, which 
uses a procedural technique to automatically generate so-called ``limited animation'' style lip sync from input
audio. While this work is related to ours, it generates lip sync with only 3 mouth shapes 
(closed, partly open, and open lip). In contrast, our approach supports a 12-viseme set that 
is typical for most modern 2D animation styles. In addition, Voice Animator runs on pre-recorded (offline)
audio.

Despite these differences in the goals and requirements of previous published lip sync methods, 
recent model-driven techniques for generating
realistic lip sync have shown the promise of learning speech-to-animation
mappings from data. In particular, the data-driven method of Taylor et al.~
\shortcite{taylor2017deep} suggests that neural networks can successfully encode the
relationships between speech (represented as phones sequences) and mouth
motions. Our work explores how we can use a recurrent network that takes
advantage of temporal context to achieve high-quality live 2D lip sync. 

\subsection*{Speech Analysis}

Our goal of converting raw audio input into a discrete sequence of 
(viseme) labels is related to classical speech analysis problems like STT
or automatic speech recognition (ASR). 
For such applications, recurrent neural networks (primarily in the form of
LSTMs) have proven very successful \cite{graves2013hybrid,yu2014automatic,graves2013speech}. In our approach, we use a basic LSTM
architecture, which allows our model to leverage temporal context from
the input audio stream to predict output visemes.
However, the low-latency requirements of our target application require a
different LSTM configuration than many STT or ASR models. In particular, we
cannot rely on any significant amount of future information, which precludes
the use of bidirectional LSTMs~\cite{graves2014towards,fan2015photo}. 
%
In addition, the lack of existing large corpora
of hand-animated 2D lip sync data (and the high cost of collecting such data)
means that we cannot rely on training sets with many hours of data, which is the
typical amount used to train most STT and ASR models.
On the other hand, our output domain (a small set of viseme classes) is much
more constrained than STT or ASR. By leveraging the restricted nature
of our problem, we achieve a low-latency model that requires a modest amount of
data to train.

\newcommand{\xa}[0]{$x_{\text{audio}}$}
\newcommand{\SX}[0]{\emph{SX}}
\newcommand{\SXtrain}[0]
{\emph{SX}$_\text{train}$}

\newcommand{\SI}[0]{\emph{SI}}
\newcommand{\SItrain}[0]
{\emph{SI}$_\text{train}$}

\newcommand{\SA}[0]{\emph{SA}}
\newcommand{\SAtrain}[0]
{\emph{SA}$_\text{train}$}

\newcommand{\rref}[0]{$R_{\text{ref}}$}
\newcommand{\visref}[0]{$V(R_{\text{ref}})$}

\begin{figure}
  \centering
  \includegraphics[width=\columnwidth]{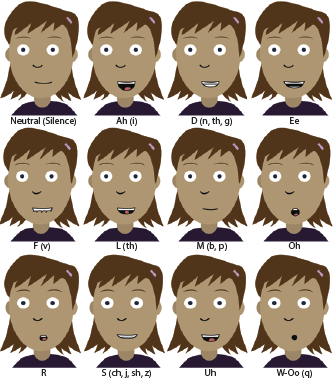}

  \caption{Chloe's Viseme Set. Additional associated sounds in parentheses.}

  \label{fig:visemes}
\end{figure}

\section{Approach}

We formulate the problem of live 2D lip sync as follows. Given a
continuous stream of audio samples representing the input speech, the goal
is to automatically output a corresponding sequence of visemes. We use the 
12 viseme classes defined by Ch (see Figure~\ref{fig:visemes}), which is similar
to other standard viseme sets in both commercial tools (e.g., ToonBoom \cite{toonboom}, CrazyTalk \cite{crazytalk} and previous research~\cite{edwards2016jali,ezzat1998miketalk,cappelletta2012phoneme,lei2003context}. 

In addition to being accurate, the technique must satisfy two main
requirements. First, the method must be fast enough to support live applications. As with any
real-time audio processing pipeline, there will necessarily be some latency in the lip sync computation. For instance, simply converting audio samples into
standard features typically requires frequency analysis on temporal windows of samples. To prevent viseme changes from appearing ``late'' with respect to the
speech, live animation broadcasts often delay the audio slightly. The size of the delay must be large enough to produce a good audio-visual alignment where
viseme changes occur simultaneously with the audio changes. In fact, some
animation literature suggests timing viseme transitions slightly early (1--2
frames at 24fps) with respect to the audio~\cite{thomas1995illusion}. At the
same time, the delay must be small enough to enable natural interactions with other
actors and the audience without awkward pauses in the animated character's
responses. We consulted with several live animation production teams and found that 
200--300ms is a reasonable target for live lip sync latench;
e.g., the live Simpsons broadcast delayed Homer's voice by 500ms~\cite{LiveSimpsonsBlog}
and livestreams often use a 150--200ms audio delay.

The second requirement involves training data. As noted earlier, data-driven
methods have proven very successful for various speech analysis problems.
However, supervised training data (i.e., hand-authored viseme sequences) is extremely time-intensive to create; we obtained quotes from professional
animators estimating five to seven hours of animation work to lip sync each minute of speech. As a result, it is difficult to obtain very large training corpora of hand-animated results. For example, collecting the equivalent amount of training data used by other recent
audio-driven models like Suwajanakorn et al.~\shortcite{suwajanakorn2017synthesizing} (17
hours) and Taylor et al.~\shortcite{taylor2017deep} (8 hours) would be extremely
costly. We aim for a method that requires an order of magnitude fewer
data.

Given these requirements, we developed a machine learning approach that
generates live 2D lip sync with less than 200ms latency using
13--20 minutes of hand-animated training data. We leverage a compact
recurrent model with relatively few parameters that incorporates a small but
useful amount of lookahead in both the input feature descriptor and the
configuration of the model itself. We also describe a simple data augmentation
scheme that leverages the inherent structure of the TIMIT speech
dataset to amplify hand-animated viseme sequences by a factor of four. The following sections describe our proposed model and training procedure.

\subsection{Model}

\begin{figure}
  \centering
  \includegraphics[width=0.95\columnwidth]{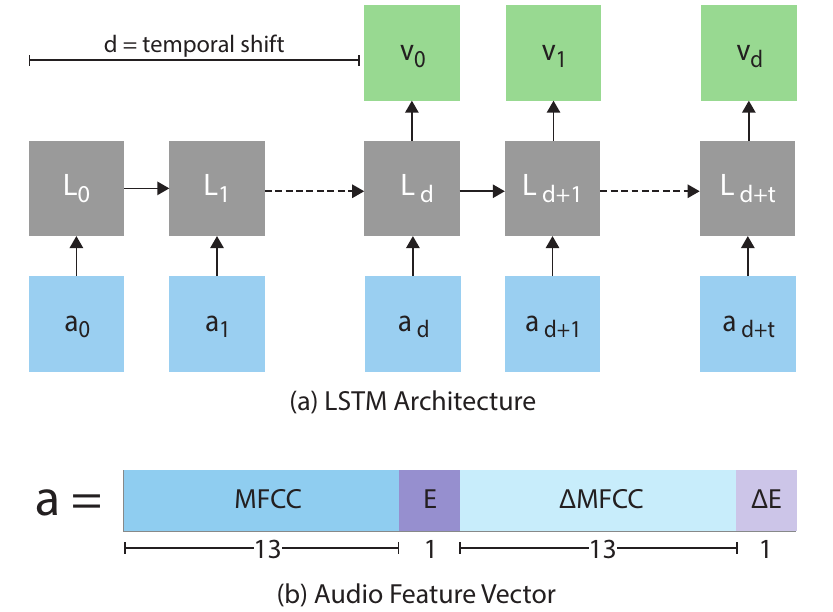}
  \caption{Lip Sync Model. We use a unidirectional single-layer LSTM with a
  temporal shift $d$ of 6 feature vectors (60ms) (a). The audio feature
  $a$ consists of MFCC, log mean energy, and their first temporal derivatives 
  (b).}
  \label{fig:lstm}
\end{figure}

Based on the success of recurrent neural networks in many speech analysis
applications, we adopt an LSTM architecture for our problem. Our
model takes in
a sequence of feature vectors \small $(a_0,a_1,\cdots,a_N)$
\normalsize derived from
streaming audio and outputs a
corresponding sequence of visemes \small $(v_0,v_1,\cdots,v_N)$ \normalsize
(see Figure~
\ref{fig:lstm}a). The latency
restrictions of our
application preclude the use of a bidirectional LSTM. Thus, we use a standard
unidirectional single-layer LSTM with a 200-dimensional hidden state that is
mapped linearly to 12 output viseme classes. The viseme with
the maximum score is the model prediction. We note that our initial experiments
explored the use of Hidden Markov Models (HMMs) to convert audio observations
into visemes, but we found it challenging to pre-define a hidden state space
that captures the appropriate amount of temporal context.
While the overall configuration of our LSTM does not deviate significantly from
previous work, there are a few specific design decisions that were important for
getting the model to perform well.

\subsubsection*{Feature Representation}

While it is possible to train a model that operates directly on raw audio
samples, most speech analysis applications use mel-frequency cepstrum
coefficients (MFCCs) \cite{walker2004sphinx} as the input feature representation. 
MFCCs are a frequency-based representation with non-linearly spaced frequency bands that roughly match
the response of the human auditory system. In our pipeline, we process the input
audio stream by computing MFCCs (with 13 coefficients) on a sliding 25ms window
with a stride of 10ms (i.e., at 100Hz), which is a typical setup for many speech
processing techniques. Before computing MFCCs, we compress and boost the input
audio levels using the online Hard Limiter filter in Adobe Audition, which runs
in real-time.

In addition to the raw MFCC values, some previous methods concatenate
derivatives of the coefficients to the feature representation~\cite{eronen2006audio,graves2005framewise,suwajanakorn2017synthesizing}.  Such derivatives are particularly important for our
application because viseme transitions often correlate with audio changes that
in turn cause large MFCC changes. One challenge with such derivatives is that
they can be noisy if computed at the same 100Hz frequency as the MFCCs
themselves. A standard solution is to average derivatives over a larger temporal
region, which sacrifices latency for smoother derivatives. We found that
estimating derivatives using averaged finite differences between MFCCs computed
two windows before and after the current MFCC window provides a good tradeoff
for our application. An additional benefit of this
derivative computation is that it provides the model with a small amount of
lookahead since each feature vector incorporates information from two MFCC
windows into the future.

In our experiments, we found that the energy of the audio signal can sometimes
be a useful descriptor as well. Thus, we add the log-energy and its derivative as
two additional scalars to form a 28-dimensional feature (see Figure~\ref{fig:lstm}b).

\subsubsection*{Temporal Shift}

Since LSTMs can make use of history, our model has the ability to learn how
animators map a sequence of sounds to one or more visemes.  However, we found
that using past information alone was not sufficient and resulted in chattery viseme transitions. One potential reason for these
problems is that, as noted above, animators often change visemes slightly ahead
of the speech~\cite{thomas1995illusion}. Thus, depriving the model of any future
information may be eliminating important audio cues for many viseme transitions.
To address this issue, we simply shift which viseme the model predicts with
respect to the input audio sequence. In particular, for the current audio
feature vector $x_t$, we predict the viseme that appears $d$ windows in
the past
at $x_{t-d}$ (see Figure~\ref{fig:lstm}a). In other words, the model has access
to $d$ future feature vectors
when predicting a viseme. We found that $d = 6$ provides sufficient
lookahead. Adding this future context does not require any
modifications to the network architecture, although it does add an additional
60ms of latency to the model.

\subsubsection*{Filtering}

Our model outputs viseme predictions at 100Hz. For live animation,
the target frame rate is typically 24fps. We apply two types of filtering to
convert the 100Hz output to 24fps.\\ 
\\
{\bf Removing noise from predictions.} Our model is generally able to predict
good viseme sequences. However, at 100Hz, we occasionally encounter spurious
noisy predictions.  Since these errors are typically very short in
duration, we use a small lookahead to filter them out. For any viseme prediction
that is different from the previous prediction (i.e., a viseme transition), we
consider the subsequent three predictions. If the new viseme holds across this
block, then we keep it as-is. Otherwise, we replace the new viseme with
the previous prediction. This filtering mechanism adds 30ms of latency.\\
\\
{\bf Removing 1-frame visemes.} After removing noise from the 100Hz model
predictions, we subsample to produce visemes at the target 24fps rate.
As a rule, animators never show a given viseme for less than two frames. To
enforce this constraint, we do not allow a viseme to change after a single
frame. This simple rule does not increase the latency of the system since it
just remembers the last viseme duration.


These mechanisms reduce flashing artifacts that sometimes arise when directly subsampling the 100Hz model output. 

\begin{figure}
  \centering
  \includegraphics[width=\columnwidth]{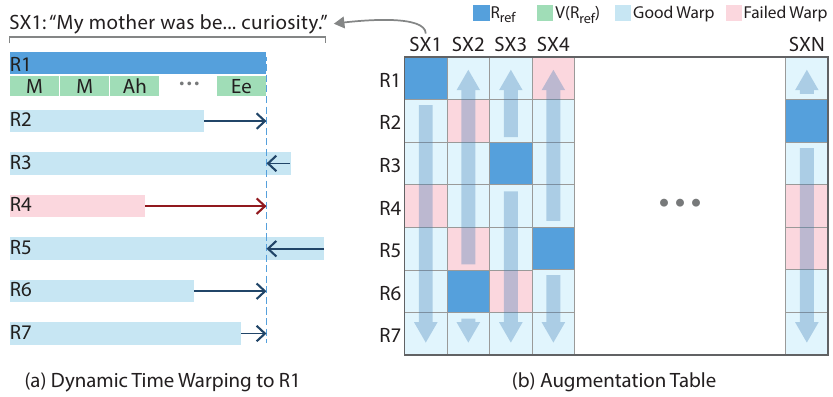}
  \caption{Data Augmentation. Each reference recording has an associated
  hand-animated viseme sequence. We automatically time warp
  other recordings of the same sentence to align with each reference recording 
  (a). This procedure allows us to create new input-output
  training pairs for every successfully warped recording.}
  \label{fig:dataAugmentation}
\end{figure}

\subsection{Training}\label{ss:training}

Training our lip sync model requires pairs of input speech recordings with
output hand-animated viseme sequences. For each input recording, we
compute the corresponding sequence of audio feature vectors, run each vector
through the network to obtain a viseme prediction, and use backpropagation
through time to optimize the model parameters. We use cross-entropy loss to
penalize classification errors with respect to the hand-animated viseme
sequence. 
The ground truth viseme sequences are animated at 24fps, so we upsample them to
match the 100Hz frequency of our model.

\subsubsection*{Data Augmentation}

In order for the model to learn the relationships between speech sounds and
visemes, the training data should cover the full spectrum of phones and common
transitions. Moreover, since we want our model to generalize to arbitrary input
voices, it is important for the training set to include a large diversity of
speakers. However, as noted above, hand-animated lip sync data is extremely
expensive to generate, which makes it difficult to collect a large collection of
input-output pairs that exhibit both phonetic and speaker diversity.

To address this problem, we leverage a simple but important insight. We do not
have to treat phonetic and speaker diversity as separate, orthogonal properties.
If we select a set of phonetically diverse sentences and record
multiple different speakers reading each sentence, then we can obtain a corpus
of speech examples that is diverse along both axes but with a useful structure
that we can exploit for data augmentation. In particular, if we manually
specify the lip sync for one speaker's recording of a given sentence, then it is
likely the case that the same sequence of visemes could be used to obtain a good
lip sync result for the other recordings of the sentence, provided that
we can align the visemes temporally to each recording. 
Fortunately, the TIMIT dataset, which has been used successfully to train many
speech analysis models, has exactly this structure. The subset of 450 unique
\SX\ sentences in TIMIT is designed to be compact and phonetically diverse,
and the corpus includes 7 recordings of each sentence by different
speakers. Overall, the recordings span 630 speakers and 8 dialects.

Based on this insight, our data augmentation works as follows. We select a
collection of reference recordings of unique \SX\ sentences and obtain the
corresponding hand-animated viseme sequences. 
For each reference recording \rref, we apply dynamic time warping~\cite{berndt1994using} to
align all other recordings of the same sentence to \rref\ (Figure~\ref {fig:dataAugmentation}a). 
We use the warping implementation in the Automatic Speech Alignment feature of
Adobe Audition. 
Since warping generally works better from male-to-male and female-to-female
voices, we only run the alignment between recordings with the same gender.
To filter out cases where the alignment fails, we discard any warped recordings
whose durations are significantly different from \rref.
Finally, we associate each \rref\ and the successfully aligned recordings with
the same hand-animated viseme sequence \visref\ to use as training pairs for our
model (Figure~\ref{fig:dataAugmentation}b).
This fully automated procedure allows us to augment our data by roughly a factor
of 4 based on the distribution of male-female speakers and the success rate
of the Automatic Speech Alignment.

\begin{figure*}[t!]
  \centering
  \includegraphics[width=\textwidth]{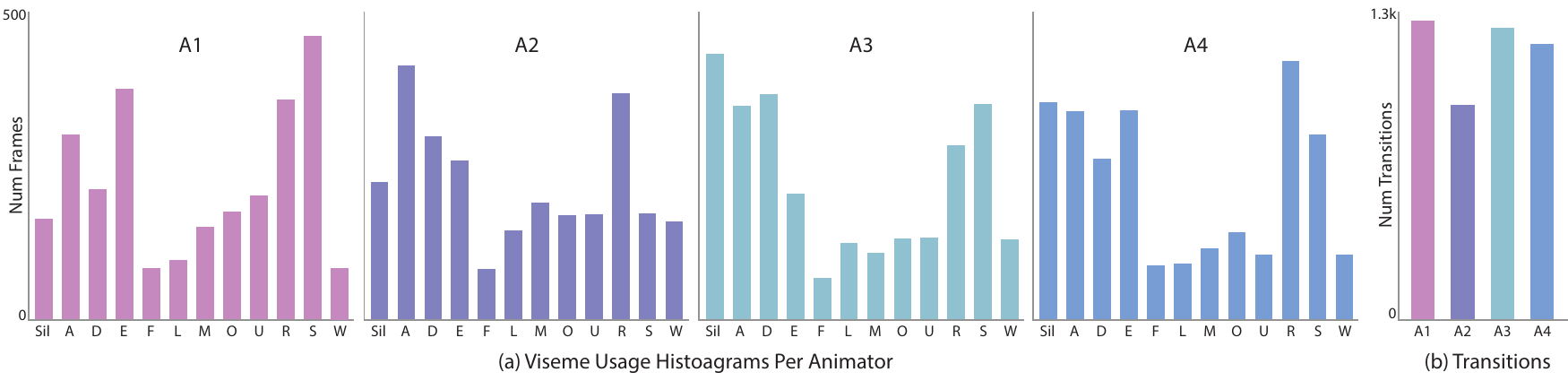}
  \caption{Analysis of Lip Sync Styles. Histograms of viseme usage (a) and raw
  transition counts (b) show that different animators prefer different visemes and
  aim for different levels of articulation.}
  \label{fig:styles}
\end{figure*}

\subsubsection*{Selecting Batches}

The TIMIT corpus consists of 450 phonetically-compact  sentences (\SX), 1890 phonetically-diverse sentences (\SI) and 2 dialect ``shibboleth'' sentences (\SA). The dataset is partitioned into training and test sets. Of the 450
\SX\ sentences, 330 sentences are in the training set \SXtrain. 
Since the \SX\ sentences are already designed to provide good
coverage of phone-to-phone transitions (with an emphasis on 
phonetic contexts that are considered difficult or particularly interesting
for speech analysis applications), we could generate our training data by
simply choosing one recording for every sentence in \SXtrain\ and obtaining a
corresponding viseme sequence.
However, we wanted to partition our training data into equivalent batches in
order to run experiments evaluating how different amounts of data affect the
performance of our model. 
To do this, we first scored all the \SXtrain\ recordings by counting the number
of distinct individual phones and phone-to-phone transitions in each
recording, using the phone transcriptions provided by the TIMIT dataset. For
each sentence, we chose the male and female recordings with the maximum scores.
Then, we generated batches of recordings by choosing subsets that include
similar distributions of high and low scoring recordings and an even mix of
male and female speakers. In the end, we produced six batches of 50 \SX\
recordings which we used for training our models. 
For our validation set, we
also created a batch of 50 recordings with a random distribution of recordings
from \SItrain, \SAtrain, and the subset of \SXtrain\ sentences not used in any
of the previously-generated six training batches. 
We obtained hand-animated viseme sequences for all seven batches.

\subsection{Model Latency}

At prediction time, the inherent latency of our model comes from the lookahead in the feature vector computation (33ms), 
the temporal shift between the input audio and output viseme predictions (60ms), and the 100Hz filtering, which takes into
account future viseme predictions (30ms). In total, this amounts to 123ms
between the time an audio sample arrives in 
the input stream and when the corresponding viseme is predicted. As noted earlier, animators sometimes show visemes slightly 
before the corresponding sounds (usually one to two frames at 24fps, or 40-80ms). The processing time required to run audio
samples through our entire pipeline, including the Hard Limiter filter before we compute feature vectors, is
1--2ms measured on a 2017 MacBook Pro laptop with a 3.1GHz Intel Core i5 processor and 8GB of memory. Thus, the total
latency in the system is approximately 165-185ms. 

\newcommand{\aone}[0]{\emph{OursA1}}
\newcommand{\atwo}[0]{\emph{OursA2}}

\section{Experiments}

We conducted several experiments to understand the behavior of our model and the
impact of our main design choices. For this quantitative analysis, we compute the per-frame accuracy of the viseme prediction at 24fps, after the filtering step in our pipeline. 

\subsection{Datasets}

We collected training data by hiring two professional animators
(A1, A2) to lip sync a set of speech recordings using Character
Animator. For consistency, they all used the default Chloe character that comes
with the application. Chloe includes the same set of 12 visemes that our model
uses (see Figure~\ref{fig:visemes}). 
We gave A1 and A2 seven batches of recordings each (six for training, one for validation), which we generated as described in Section ~\ref{ss:training}
The six training batches represented about 20 minutes of speech in total.
After propagating the hand-generated viseme sequences to the aligned \SX\
recordings using our data augmentation procedure, we obtained approximately 80 minutes of training data per animator.

To gain more insight on the differences in lip sync style, we
recruited two other animators (A3, A4) and asked all four to lip
sync an additional 27 TIMIT recordings (25 from the \SX\ recordings and 2 from the \SA\ recordings in TIMIT). These results allow us to analyze how
different animators time transitions and choose visemes for the same recordings.


\subsection{Differences in Style}

The statistics of the viseme sequences generated by the four different animators
for the same 27 recordings reveal clear differences in lip sync style. In terms
of overall viseme choices, different animators used different distributions
of visemes (Figure~\ref{fig:styles}a) and also changed visemes at different rates 
(Figure~\ref{fig:styles}b). For example, A1 and A2 use the Silent viseme far less
than A3 and A4, which suggests that they prefer sequences that do not return to
the neutral mouth pose. A1 also likes to use the S viseme much more than others.
In terms of viseme changes, A2's relatively low overall transition count suggests
that the animator prefers a smoother, less articulated style.

\subsection{Accuracy and Convergence Behavior}

We trained separate models using the full datasets that we collected from A1 (\aone) and A2 (\atwo). We used the last batch of 50 hand-animated sentences as the validation set and trained on the data from the six \SX\ batches. All the networks are trained using the Torch framework \cite{collobert2011torch7} until convergence (200 epochs) using the Adam optimizer \cite{kingma2014adam}, with a dropout ratio of 0.5 for regularization to avoid overfitting, batch size of 20, and learning rate of 0.001. On a single NVIDIA GTX-1080 GPU, training
took less than 30 minutes.
For the output layers, we used the softmax activation function for 12 viseme output classification and the cross-entropy error function to compute the classification accuracy. The per-frame viseme prediction accuracy for \aone\ is 64.37\% and \atwo\ is 66.84\%.


\subsection{Impact of Lookahead}

To evaluate the importance of using future information (albeit a small amount)
in our approach, we trained a version of \atwo\ with no temporal shift
between observations (feature vectors) and predictions (visemes) and modified
the feature vector to include derivatives computed using past MFCC windows only.
The per-frame accuracy for the no-lookahead version of \atwo\ is 59.27\%, which is significantly lower than the accuracy of \atwo\ (66.84\%) which is trained with temporal shift (d=6) and using two future windows for MFCC derivative computation. From a qualitative perspective, we notice that the model without lookahead appears to be chattery, with extra transitions around the expected viseme changes.

\begin{figure}[t!]
  \centering
  \includegraphics[width=0.95\columnwidth]{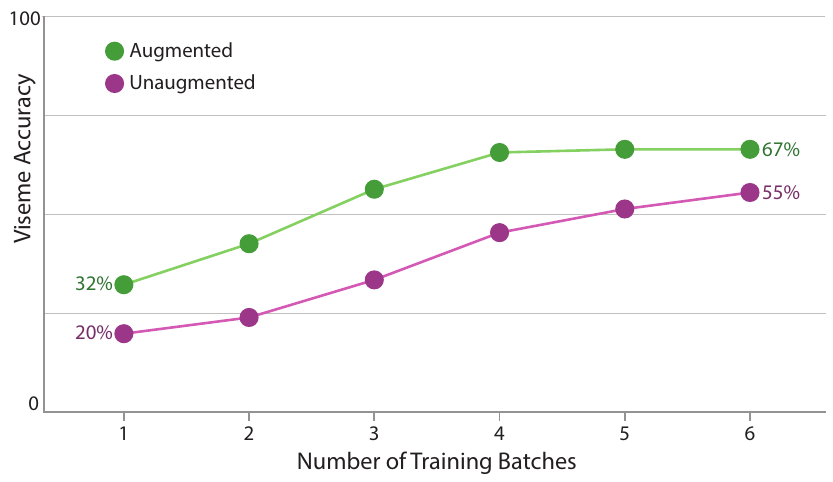}
  \caption{Impact of Data Augmentation. Augmenting the data results in a
  significant increase in accuracy, with diminishing returns after four
  augmented batches.}
  \label{fig:accuracyVsData}
\end{figure}

\subsection{Impact of LSTM context}

One advantage of using an LSTM over a non-recurrent network (e.g., the sliding window CNN of Taylor et al.~\cite{taylor2017deep}),
is that LSTMs can leverage a larger amount of (past) context without increasing the size of the feature vector. While longer feature
vectors can cover more past context, they result in larger networks that in turn require more data to train. 
To investigate how much context our model actually uses for viseme prediction, we trained different versions of \atwo\ with data that artificially
limits the amount of context the LSTM can leverage. Our initial experiments showed that the model performance does not improve with more than one second
of context, so we segmented the A2 training data into uniform chunks of several durations (200ms, 400ms, 600ms, 800ms, 1sec) and trained our LSTM on
each of these five datasets.
The per-frame viseme prediction accuracies are 24.63\%(200ms), 37.08\%(400ms), 56.44\%(600ms),	59.72\%(800ms) and 64.81\%(1sec). 
The significant increase in accuracy around 600ms suggests that our model is mainly using around 600--800ms of context, which corresponds to 60--80 
MFCC windows. In other words, these results suggest that a non-recurrent model may need to use much longer feature vectors (and thus, much more training data) to achieve comparable viseme prediction results. 

\subsection{Impact of Data Augmentation}\label{ss:augmentationExperiments}

Finally, we investigate the effect of our data augmentation technique by
training versions of \atwo\ with various amounts of data.
Specifically, we consider an unaugmented dataset that only has the hand-animated
viseme sequences, and our full augmented dataset.
We divide the A2 training data into increasing subsets of the 6 hand-animated batches 
and train the model on both the unaugmented and augmented subsets. As expected, our data
augmentation allows us to achieve much higher accuracy for the same amount of
animator work (see Figure~\ref{fig:accuracyVsData}). Moreover, there is a
clear elbow in the accuracy for the augmented data at around 4 batches, which
corresponds to roughly 13 minutes of hand-animated lip sync. In other words, an
animator may only need to provide this amount of data to train a new version of
our lip sync model. We further validate this claim in Results Section with human judgement experiments that compare the full model with the version
trained using 4 augmented batches.

\section{Results}\label{s:results}

To evaluate the quality of our live lip sync output, we collected human
judgements comparing our results against several baselines, including competing
methods, hand-animated lip sync, and different variations of our model. In informal pilot studies, we saw a slight
preference for A2's lip sync style over A1, so we used the \atwo\ results for
these comparisons.
We also conducted a small preliminary study comparing the
stylistic differences between \aone\ and \atwo\ results.

\begin{figure}[t]
  \centering
  \includegraphics[width=\columnwidth]{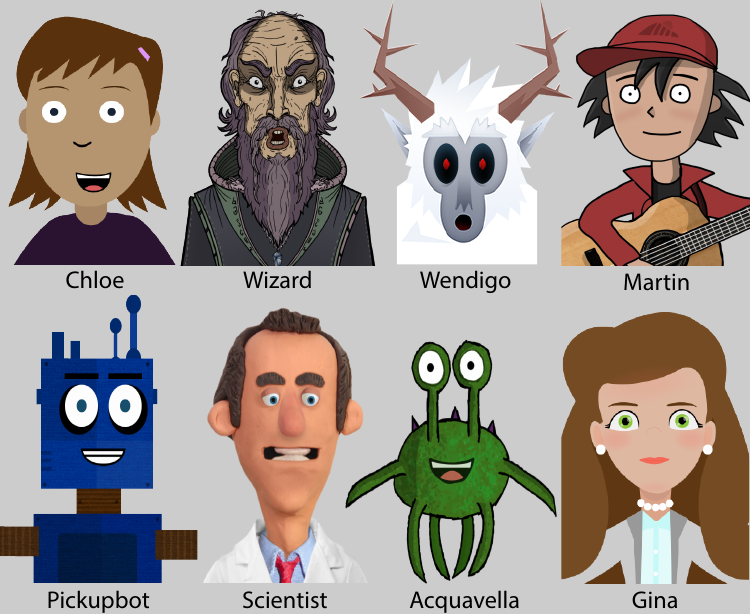}
  \caption{Characters. We used Chloe and the Wizard for our human
  judgement experiments, and we show lip sync results with the other
  characters in our video summary and supplemental materials.}
  \label{fig:characters}
\end{figure}

In addition to these comparisons, we applied our lip sync model to several
different 2D characters (see
Figure~\ref{fig:characters}) that come bundled with Character Animator.
Our video summary and supplemental materials (GitHub link: \url{https://github.com/deepalianeja/CharacterLipSync2D})
show representative lip sync results using these characters. We also include
real-time recordings that shows the system running live in a modified version of
Ch. For these recordings, we delay the audio track by 200ms to account for 
the latency of our model. As noted earlier, this type of audio delay is standard 
practice for live animation broadcasts. 

\subsection{Comparisons with Competing Methods}

We are not aware of any previous research efforts that directly support 2D
(discrete viseme) lip sync for live animation. Thus, we compared our method
against existing commercial systems. The predominant tool for live 2D
animation (including live lip sync) is Character Animator (Ch), which was used
for the live Simpsons episode, the recurring live animation segments on The Late
Show, and to our knowledge, all of the recent live animated chat sessions on
Facebook and YouTube. In addition to live lip sync, Ch also includes a higher
quality offline lip sync feature. For traditional non-live cartoon animation,
ToonBoom (TB) is an industry standard tool that also provides offline lip sync.
We compared our results using A2's model (\full) against the Ch online lip sync 
(\chon), and the offline output from both Ch (\choff) and TB (\tboff).

\subsubsection*{Procedure}

To compare our model against any one of the competing methods, we selected a
test dataset of recordings, and for each one we generated a pair of viseme
sequences using the two lip sync algorithms. We applied the lip sync to two
characters (Chloe and the Wizard, shown in Figure~
\ref{fig:characters}) that are drawn in distinct styles with visemes that look
very different. 
For each character, we 
presented pairs of lip sync results to users and asked which one they
prefer. 
We used Amazon Mechanical Turk (AMT) to collect these judgements. Based
on
pilot studies, we found that showing the lip sync results side-by-side with
separate play controls made it easy for users to review and
compare the output. Since our method uses the same set of
visemes as Ch, we were able to generate direct comparisons between our model and
both the online and offline Ch algorithms. TB uses a smaller set of eight
visemes for their automatic lip sync. To generate comparable results, we mapped
a subset of our viseme classes (S->D, L->D, Uh->Ah, R->W-Oo) to the TB visemes based on TB's published phone-to-viseme guide and then used this subset to generate lip sync from TB. For our model,
we mapped each viseme that is not in the TB subset to one of the TB visemes and
used this mapping to project our lip sync output to the TB subset.

\subsubsection*{Test Set} 

For our test dataset, we randomly chose 25 recordings from the TIMIT test set,
using the same criteria as our training batch selection process to ensure even
coverage of phones and transitions. 
To increase the diversity of our test set, we composed an additional 10
phonetically diverse sentences and recorded a man, woman and child reading each
one. We also recorded a voice actor reading each sentence in a stylized cartoon
voice. We randomly chose 25 of these non-TIMIT recordings for testing.
All test recordings were between 3--4 seconds. 
None of these recordings were used for training.
We used the same test set and procedure for all the comparisons described in the following sections.

%

\subsubsection*{Findings}

\begin{figure}[t]
  \centering
  \includegraphics[width=\columnwidth]{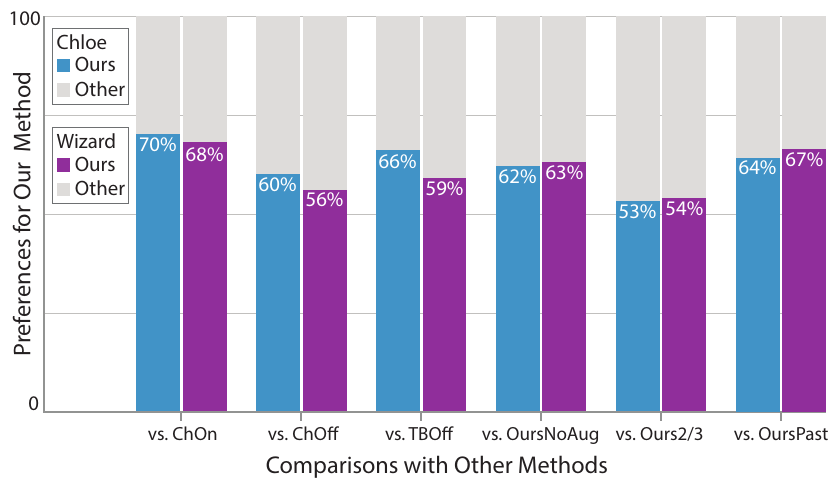}
  \caption{Human Judgements. Our method was significantly preferred over all
  commerical tools, including offline methods. Our full model was
  also preferred over versions trained with no augmented data (\noaug) and two
  thirds of
  the augmented data (\some). However, the preference over \some\ was quite
  small, which suggests that this amount of data may be sufficient to
  train an effective model.}  
  \label{fig:mturkComparisons}
\end{figure}

We collected 20 judgements for every recording (10 for each puppet), which 
resulted in 1000 judgements for each competing method. 
The left side of Figure~\ref{fig:mturkComparisons} summarizes the results of the
comparisons with Ch and TB. Our lip sync was preferred in all cases, and these
differences were statistically significant (at $p = 0.05$) based on the Binomial
Test. We are especially encouraged that our results outperformed even the
offline Ch and TB methods, which do not support live animation. Moreover, we did
not see much difference between the
results for the non-TIMIT versus TIMIT test recordings, which suggests that
our model generalizes to a broader spectrum of speakers. 
Qualitatively, we found the \choff\ and \tboff\ results to
be overly smooth (i.e., missing transitions) in many cases, while the
\chon\ output tends to be more chattery. We also saw a few cases where
\tboff\ uses visemes that clearly do not match the corresponding sound.
Our video summary shows several direct comparisons that highlight these differences. 

\subsection{Comparisons with Groundtruth}

To get a sense for how artist-generated lip sync compares to automatic results,
we compared the groundtruth (i.e., hand-animated) version of our test set against our full model (\full) and all the competing methods. 

\subsubsection*{Findings}

As expected, all the automatic methods (including ours) are preferred much less than the groundtruth: \full=13.5\%, \chon=6.1\%, \choff=13\%, \tboff=9.1\%, averaged across both Chloe and the Wizard.
While these results clearly show there is room for improvement, they also align with Figure~\ref{fig:mturkComparisons} in that our model does much better than \chon\ and somewhat better than the two competing offline methods. 

\subsection{Comparisons with Different Model Variations}

Our data augmentation experiments (Impact of Data Augmentation Section)
suggest that our model should already perform well using just four out of the
six hand-animated training batches.
To validate this conjecture, we compared the output of our full model (\full)
against a version trained with four augmented batches of hand-animated data 
(\some). As a baseline, we also compared \full\ with a model trained on all six
batches \emph{without} data augmentation (\noaug). Similarly, we compared the output
of our no-lookahead model (\past) to \full\ in order to validate the impact of lookahead 
on the perceived quality of the resulting lip sync.

\subsubsection*{Findings}
The right side of Figure~\ref{fig:mturkComparisons} shows the comparison results for the different versions of our model.
Not surprisingly, \full\ is clearly preferred over \noaug. The lack of augmented data results in lip sync with
both incorrect viseme choices and a combination of missing and extraneous
transitions. On the other hand, the preferences between \full\ and \some\ are
much more balanced, which suggests that we may only require about four batches 
(13 minutes) of hand-animated data to train an effective live lip sync model. \full\ was also distinctly 
preferred over \past\, showing the benefit of the small amount of lookahead in our full model.

\subsection{Matching Animator Styles}

While most high quality lip sync shares many characteristics, there are some
stylistic differences across different animators, as noted earlier. 
To investigate how well our approach captures the style of the training
data, we conducted a small experiment comparing the outputs of \aone\ and \atwo\ .
We randomly chose 19 hand-animated viseme sequences from each
animator that were not part of the training sets for the two models. For each
hand-animated result, we generated lip sync output from \aone\ and
\atwo\ using the corresponding speech recording and then presented the two
automatic results to the animator along with their
own hand-animated sequence as a reference. We then asked the animator to pick
which of the model-generated results most resembled the reference. 
We used the Chloe character for this experiment.

\subsubsection*{Findings}

Each animator chose the ``correct'' result (i.e., the one generated by the
model trained on their own lip sync data) more often than the alternative. A1
chose correctly in 12/19 and A2 chose correctly in 15/19 comparisons. 
While these are far from conclusive results, they suggest that our model is
able to learn at least some of the characteristics that distinguish different
lip sync styles. 

\subsection{Impact on Performers}

The experiments described above evaluate the quality of lip sync that our model outputs as judged by
people who are viewing the animation. We also wanted to gather feedback on how our lip sync
techniques affect \emph{performers} who are controlling live animated characters with their voices. 
In particular, we wondered whether the improved quality of our lip sync or the small amount of latency in our model would 
have an impact (positive or negative) on perfomers. To this end, we conducted a small user study with nine participants comparing 
three lip sync algorithms: \chon, \past, and \full. To minimize the differences between the conditions, we implemented \past\ and \full\ within Ch. 
We used a within-subject design where each participant used all three conditions (with the order counterbalanced via a 3x3 Latin square) to control 
Chloe's mouth movements. To simulate a live animation setting, we asked each participant to answer 6 questions (two per condition) as if they were being interviewed as Chloe. During the performance, we used the relevant lip sync method to show the participant live feedback of Chloe's mouth being animated. At the end of the session, we asked participants to rate each condition based on the effectiveness of the live feedback, on a scale from 1 (very distracting) to 5 (very useful) . We also solicited freeform comments on the task. Each session lasted roughly 20 minutes.

\subsubsection*{Findings}
\setlength{\columnsep}{3pt}
\setlength{\intextsep}{0.25pt}
\begin{wrapfigure}{r}{1.7in}
  \begin{center}
    \includegraphics[width=1.65in]{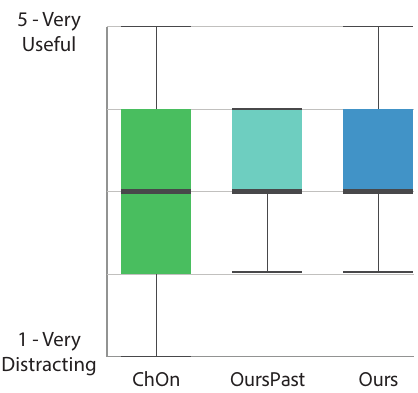}
  \end{center}
\end{wrapfigure}
We summarize the collected ratings for each condition using a box and whisker plot, as shown on the right. The data does not show any discernible 
difference in how participants rated the usefulness of the live feedback across the different algorithms. 
In particular, the latency of our full model did not have a noticeable negative impact on the performers. The comments from participants suggest that the cognitive load of performing (e.g., thinking of how to best answer a question) makes it hard to focus on the details of the live feedback. In other words, the results of this study suggest that the quality of live lip sync is mainly relevant for viewers (as shown in our human judgement experiments) rather than performers.

\section{Applications}

Our work supports a wide range of emerging live animation usage scenarios. For example, animated characters can interact directly with live audiences on social media via text-based chat. Another application is for multiple performers to control different characters who can respond and react to each other live within the same scene. In this setting, the performers do not even have to be physically co-located. Yet another use case is for live actors to interact with animated characters in hybrid scenes.
Across these applications, high-quality live 2D lip sync helps create a convincing connection between the performer(s) and the audience. 
We demonstrate all of these scenarios in our submission video using our full lip sync model implemented within Ch. For the hybrid scene, we used Open Broadcaster Software~\cite{obs} to composite the animated character into the live video. 

While our approach was motivated by common 2D animation styles that use discrete viseme sets, our method also applies to some 3D styles. For instance, discrete visemes are sometimes used with rendered 3D animation to create stylized lip sync (e.g., the recent Lego movies, Bubble Guppies on Nickelodeon). Stop motion is another form of 3D animation where visemes are often used. In this case, physical mouth shapes are photographed and then composited on the character's face while they talk. Our lip sync model can be applied to create live animations for these types of 3D characters. As an example, our submission video includes segments with the stop motion Scientist character shown in Figure~\ref{fig:characters}.

\section{Limitations}

There are two main limitations with our current method that stem from our source
of training data. The TIMIT recordings all contain clean, high-quality audio of
spoken sentences. As a result, our model performs best on input with similar
characteristics. While this is fine for most usage scenarios, there are
situations where the input audio may contain background noise or distortions due
to the recording environment or microphone quality. For example, capturing
speech with the onboard microphone of a laptop in an open room produces
noticeably lower quality lip sync output than using even a decent quality USB
microphone in a reasonably insulated space. Note that the production teams for
almost all live broadcasts already have access to high end microphones and sound
booths, which eliminates this problem. In addition, we noticed that
vocal input that is very different from conversational speech (e.g., singing,
where vowels are often held for long durations) also produces suboptimal
results.

We do not believe these are fundamental limitations of our approach. For
example, we could potentially collect more training data or, better yet,
employ additional data augmentation techniques to help the model learn how to
better handle a wider range of audio input. To support singing, we may also need
to include slightly different audio features. Of course, we would need to
conduct additional experiments to confirm these conjectures. 

\vspace{-0.2mm}
\section{Conclusions and Future Work}

Our work addresses a key technical challenge in the emerging domain of
live 2D animation. Accurate, low-latency lip sync is critical for almost all
live animation settings, and our extensive human judgement experiments
demonstrate that our technique improves upon existing state-of-the-art 2D lip
sync engines, most of which require offline processing. Thus, we believe our work has immediate practical implications
for both live and even non-live 2D animation production. Moreover, we are not aware of previous 2D lip sync work with 
similarly comprehensive comparisons against commercial tools. To aid future research, we will share the artwork and 
viseme sequences from our human judgement experiments.

We see many exciting opportunities for future work:\\
\\
{\bf Fine Tuning for Style.} While our data augmentation strategy reduces the
training data requirements, hand-animating enough lip sync to train a
new model still requires a significant amount of work. It is possible that we do
not need to retrain the entire model from scratch for every new lip sync style.
It would be interesting to explore various fine-tuning strategies that would
allow animators to adapt the model to different styles with a much smaller
amount of user input.\\
\\
{\bf Tunable Parameters.} A related idea is to directly learn a lip sync model
that explicitly includes tunable stylistic parameters. While this may require a
much larger training dataset, the potential benefit is a model that is general
enough to support a range of lip sync styles without additional training.\\
\\
{\bf Perceptual Differences in Lip Sync.} In our experiments, we observed that
the simple cross-entropy loss we use to train our model does not accurately
reflect the most relevant perceptual differences between lip sync sequences. In
particular, certain discrepancies (e.g., missing a transition or replacing a
closed mouth viseme with an open mouth viseme) are much more obvious and
objectionable than others. Designing or learning a perceptually-based loss may
lead to improvements in the resulting model.\\
\\
{\bf Machine Learning for 2D Animation.} Our work demonstrates a way to encode
artistic rules for 2D lip sync with recurrant neural networks.
We believe there are many more opportunities to apply modern
machine learning techniques to improve 2D animation workflows. Thus far, one
challenge for this domain has been the paucity of training data, which is
expensive to collect. However, as we show in this paper, there may be ways to
leverage structured data and automatic editing algorithms (e.g., dynamic time
warping) to maximize the utility of hand-crafted animation data.

\balance{}

\bibliographystyle{SIGCHI-Reference-Format}
\bibliography{sample}

\end{document}